# EOG Communication Interface for Quadriplegics: Prototype & Signal Processing

*Aniket Raj*
*a_raj@ee.iitr.ac.in*
*Department of Electrical Engineering, IIT Roorkee*

*Amit Kumar*
*a_kumar7@ee.iitr.ac.in*
*Department of Electrical Engineering, IIT Roorkee*

## ABSTRACT

*Electrooculography (EOG) is an electrophysiological signal that determines the human eye orientation and is therefore widely used in Human Tracking Interfaces (HCI). The purpose of this project is to develop a communication method for quadriplegic patients using EOG signals aimed at text and voice generation. The system consists of 3D eye movement tracking embedded using a custom-built prototype to measure the eyeball's left-right and up-down movements.*

*The ESP32 board, which has a set of parameters to convert the data into content displayed on LCDs and MP3 players, is used to capture and process the signal. helps people by facilitating more natural and efficient symptom expression. The blink system will be able to incorporate face masks and more eye tests as it continues to develop. Even if it might work, more research and clinical trials are needed to evaluate the system's usefulness and ensure that it performs as planned in real-world scenarios. With this project, assistive technology will make significant progress and improve the lives of many who suffer from severe motor impairments.*

## 1. Introduction

Language is an important means of communication because it allows people to express their thoughts, feelings, and emotions. However, quadriplegics find it extremely difficult to communicate. People who have quadriplegia, a disorder in which all voluntary muscles are lost except for those used to move their eyes and blink, have limited ability to communicate. Even while technology has been used to provide a lot of communication services, many techniques—like voice recognition and head movements—are sometimes useless, especially in chaotic or noisy settings.

For people with severe motor impairments, electrooculography (EOG) is a promising method. EOG provides a way to communicate with HCI (human-machine interface) devices by measuring the electrical potential generated by eye movements. The aim of the project is to develop an EOG-based communication interface specifically for quadriplegic patients. The device tracks eye movements and sends users text and audio messages by tracking the corneal-retinal potential difference (EOG) signals.

Despite these efforts, many systems continue to use graphical user interfaces or suffer from problems such as signal drift and erroneous classification. The project aims to overcome these issues by creating PCBs for signal acquisition and processing, as well as algorithms that convert eye movements into text and speech. The goal is to create a better, more effective communication system that allows patients to be independent and interact with their environment.

Future improvements will include expanding sensing capabilities and adding more sensors to improve performance and reliability.

## 2. Literature Review

EOG-based systems are rapidly becoming an assistive communication technology for those suffering quadriplegia and similarly severe physical disabilities. They record eye movements to control devices, thereby giving one a non-invasive communication pathway when traditional interfaces cannot be used.

Rania A. (2024) designed a new EOG-based system in which text and voice outputs were used as modes of communication solely by the eye movements of patients who have quadriplegia [1]. It has been proved in their research work that the output may change into a lot of different types from EOG signals, which makes it easier to have communication. In the same way, Siriwadee A. (2012) focused on the designing of an eight-directional eye movement control system in which eight directions through which information can be extracted would generate more commands toward better HCI application possibilities [2].

However, Belkhiria C's work holds that signal processing of EOG has to be effective in 2022 since a strong filtering for the elimination of noise dominates the challenges surrounding the acquisition of the EOG signals. Their work underscores the requirement of clear signal processing techniques for reliable performance especially in a real-time application where accuracy is of the essence. Siriwadee A. In addition, here is an extended discussion on the structured approaches for EOG signal interpretation in triggering specific commands in biomedical applications, further explaining the importance of accurate control [4]. Recent work from Lourens M. (2022), through statistical analysis of the variables, has further elaborated the concept, and it states that ML can estimate the stress condition based on physiological signals and it can also provide scope for early intervention in stress management. [5].

The same direction is Rim B. Refers to (2020) sensor technologies and embedded systems find development for acquiring physiological signals with emphasis on the energy efficiency as well as signal transmission accuracy, which also raise features in the proposed design here [6]. Application of machine learning on the classification of physiological signals-Luri Crac. (2022) concluded that these algorithms improve the interpretation of EOG for better text and voice output [7].

The studies just discussed constitute a good basis for our EOG-based communication interface. Advanced filtering, adaptive algorithms and efficient sensor technology used in this project make the interface robust, non-invasive, and cost-effective, hence offering an alternative with regard to optimization of the system with machine learning.

## 3. Methodology

The process of designing and implementing the EOG-based communication system encompasses many essential stages, such as hardware development, signal acquisition processing, and output.

### 3.1 Hardware Design

- **Microcontroller and PCB design:** The Arduino Ide programming environment was compatible with the XIAESP32 C3 microcontroller, which was chosen for its performance. For data acquisition, a specially designed PCB with an INA128UA instrumentation amplifier, TL084 operational amplifier, and LM741 operational amplifier was made. This combination of precision amplifiers and filters ensures accurate, stable, and efficient signal processing for sensor data in this prototype, which proves to be very useful.

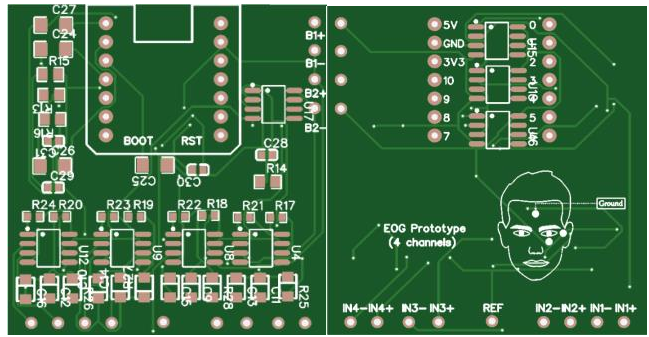
Fig. PCB Layout (a) Top Layer (b) Bottom Layer

- **Electrode Placement:** Electrodes are positioned strategically around the eyes to capture left-right and up-down movements to the right and left of the eyes for horizontal movements. These electrodes detect electrical signals generated by the eye muscles.

**3.2 Signal Acquisition and Processing**

- **Amplification and Filtering:** Firstly, the EOG signals coming out from the electrodes are amplified using the INA128UA instrumentation amplifier, which has a gain of 1. The instrumentation amplifier INA128 is selected for its enhanced specifications, which are crucial in the accuracy of the system, low noise, high Common-Mode Rejection Ratio (CMRR), and wide supply range. Further, a Bandpass Filter with a gain of 152.5 is included, and it is designed to operate within a frequency range of 20 Hz to 500 Hz. The filter consists of resistors, capacitors, and the LM741 operational amplifier, which is effective due to the high slew rate, low offset voltage, low power consumption, low noise, output short circuit protection, low total harmonic distortion, wide supply voltage.
- **Data processing:** XIAESP32 C3 receives the filtered signal. Based on specified data, ESP32 determines the kind of eye movement that takes place. A directive like "Call a doctor" will appear on the LCD screen when a certain signal is identified.

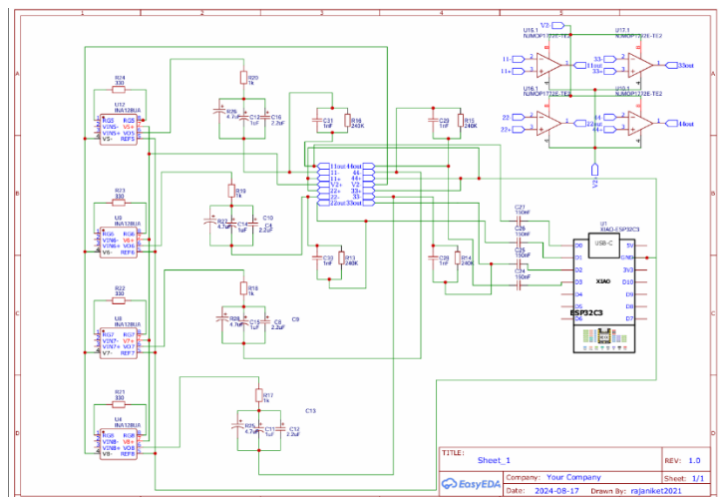
Fig. Circuit of Prototype

### 3.3 Output Generation

- **Text and Voice Outputs:** The voice output is provided with the help of the DFPlayer Mini. It delivers messages recorded previously via a speaker, and the audio files are embedded in SanDisk SD cards. Because LCDs are used for text outputs, a dual mode communication system with audio and visual feedback is created.

### 3.4 Calibration and Validation

- **Calibration:** The eye movement ranges of each user are calibrated because the potential difference varies depending on the individual's expected degrees of gaze. This involves ascertaining the lowest and highest level of movement towards left, right, up, and down directions, with the raw EOG data being analysed to ascribe layers of these levels.

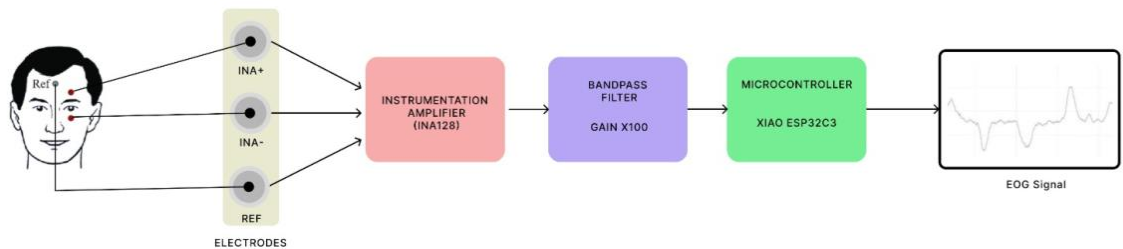

Fig. Signal Acquisition and Processing

## 4. Working

The EOG interfaces the various hardware and software components in the form of signal capture, data analysis, and output generation.

### 4.1 Signal Capture

- **Electrode Placement:** Several Electrodes are required to be applied on the skin surface near the eyes so as to capture movements in four directions viz, right, left, up, down. The PCB is designed such that these signals can get transformed into analogue signals.
- **Signal Transmission:** Now the raw signal of EOG will be transmitted to XIAESP32 C3 so that further processing may be done.

### 4.2 Data Processing

- **Signal Interpretation:** The XIAESP32 C3 uses algorithms in Arduino IDE, which decode the EOG signals. The application loops continually and compares all incoming signals against some default ranges for up, down, left, and right eye movements. The senses applied for detecting the eye's direction use the if conditions.

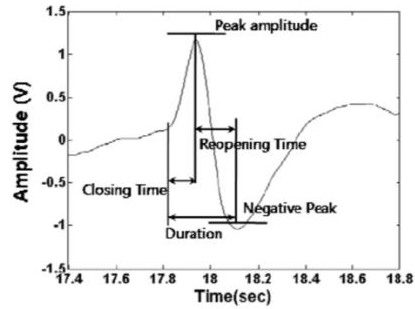
Fig. Interpretation of acquired Signal [8]

- **Calibration and Matching:** The system is calibrated to define specific signal ranges for each movement. Real-time data is compared by ESP32 with these calibrated ranges to define the corresponding command. The acquired signals are filtered using 4th order Butterworth Filter.

**Overview of Filter Design and Biquad Implementation**
This is the code for a 4th-order Butterworth low-pass filter. The Butterworth filters are designed with maximally flat response in the passband for which no ripples are introduced and there will be a smooth roll-off at the cutoff edge of the passband. This one here is implemented as a cascade of second-order sections, often known as biquads.

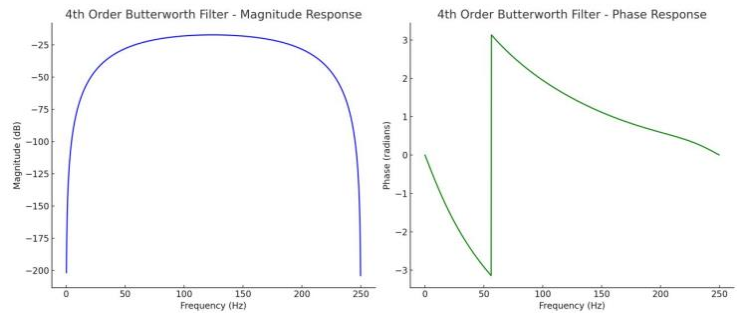
Fig. Analysis of EOG signals

**Why Second-Order Sections?**
High-order filters, in this case, are often broken into biquads for better numerical stability and lower rounding errors generally associated with filtering in real-time embedded systems.
**Step 1: Understanding Each Biquad Section**
Each biquad (second-order section) follows this difference equation for digital filtering:
>   $y[n] = b_0 \cdot x[n] + b_1 \cdot x[n-1] + b_2 \cdot x[n-2] - a_1 \cdot y[n-1] - a_2 \cdot y[n-2]$
>   Where:
>   x[n]: Current input sample
>   x[n−1] and x[n−2]: Previous input samples
>   y[n]: Current output sample
>   y[n−1] and y[n−2]: Previous output samples
>   b0, b1, b2: Feedforward coefficients
>   a1, a2: Feedback coefficients

This equation represents how a filter is recursive, meaning that every sample of the output depends on the current and past input values and past values of output.

**Step 2: Filter Coefficients and Frequency Response**
Every biquad section in this code has some coefficients.

**First Section Coefficients**
For the first section:
  **(i) Feedforward coefficients:**
      $b_0=0.09797471$
      $b_1=0.19594942$
      $b_2=0.09797471$
  **(ii) Feedback Coefficients:**
      $a_1=0.02977423$
      $a_2=0.04296318$

This set of coefficients introduces some attenuation at higher frequencies but let low frequencies pass through largely unaltered. Being part of a cascade, its effect adds up to the rest of the sections to produce a 4th-order response.

**Second Section Coefficients**
For the second section:
  **(i) Feedforward coefficients:** $b_0=1.0$, $b_1=2.0$, $b_2=1.0$
  **(ii) Feedback coefficients**: $a_1 = 0.08383952$, $a_2 = 0.46067709$

These coefficients help reinforce the low-pass response, providing stronger attenuation as frequencies increase.

**Third and Fourth Section Coefficients**
The third and fourth sections also use:
$b_0 =1.0$, $b_1 =-2.0$, $b_2 =1.0$,
  **(i) Feedback coefficients:**
      **Third section:** $a_1= -1.92167271$, $a_2=0.92347975$
      **Fourth section:** $a_1= -1.96758891$, $a_2=0.96933514$

These sections are designed to yield extensive attenuation in the stopband that can improve roll off of the filter overall near the cutoff frequency.

**Step 3: Computing the Filtered Output**
Within each section, it calculates an intermediate filtered output and updates the state variables $z_1$ and $z_2$, or previous input values to keep track of the filter's history for the next sample.

**The sequence of operations:**
1. Calculate the intermediate input x
      $x = output − a_1 \cdot z_1 − a_2 \cdot z_2$
2. Calculate the new output:
      $output = b_0 \cdot x + b_1 \cdot z_1 + b_2 \cdot z_2$
3. Update the state variables $z_1$ and $z_2$ for the next cycle.

**Step 4: The total response of this is a 4th-order filter.**
For each biquad, attenuation starts at the cutoff frequency and beyond, so the slope of the magnitude response is very steep. This ensures that high-frequency noise is removed with good efficacy, and the signal within the passband is preserved.

**Frequency and Phase Analysis**

To analyze this mathematically:
- The frequency response of each section can be found by applying the Z-transform to the difference equation.
- For the entire filter, the overall frequency response

$$H_{total}(e^{j\omega}) = H_1(e^{j\omega}) \cdot H_2(e^{j\omega}) \cdot H_3(e^{j\omega}) \cdot H_4(e^{j\omega})$$

The magnitude response (which we graphed) and phase response derive directly from $H_{total}(e^{jw})$, showing the low-pass characteristics and the phase shift introduced by cascading all sections.

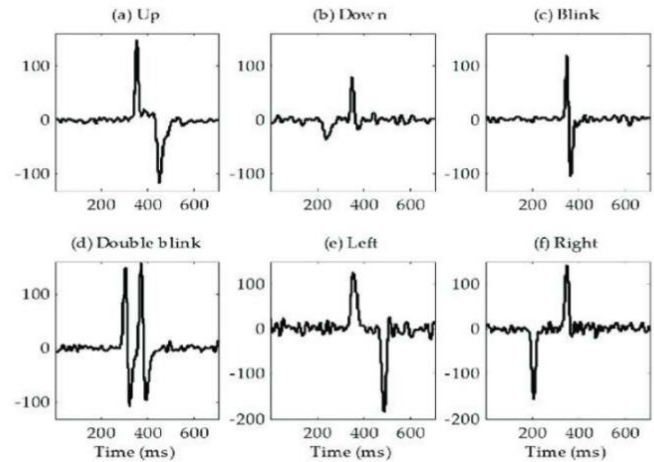

Fig. Forehead EOG Signal Patterns after signal processing (a) Up (long blink); (b) Down; (c) Blink; (d) Double Blink; (e) Left and (f) Right

## 4.3 Output Generation

- **Text Presentation:** Once the movement is sensed, the corresponding command appears on a 16x2 display. A leftward eye movement can also call "Call the Doctor".
- **Visual Indicators:** The movement detected is indicated by different color illuminations on the board. This gives instantaneous feedback to the user.
- **Voice Output:** The mini MP3 player contains pre-recorded audio files associated with the text displayed and can play them back.
- **Remote Monitoring:** The command data is communicated back to a remote monitoring application where caregivers or remote users can view the commands in real-time.

## 4.4 Control and Calibration

- **System Control:** There is control and system that is designed with a push button which would allow the system to activate or deactivate it. Ranges for the user are calibrated through eye movement.

This would allow the system to translate eye movements clearly into action that can be reacted upon as the communication of quadriplegic patients.

## 5. Future Works

In the current phase of the project, a great amount of work has been accomplished concerning the development of EOG based communication interface. The PCB is assembled and the signal acquisition is completed as shown in the fig. Analysis and synthesis of higher-order data processing algorithms and spectrum generation techniques is also done on the acquired signals. This PCB is fundamental for the accurate acquisition and processing of EOG signals. The assembly and the testing of the PCB and Signal Acquisition of PC marks a critical milestone.

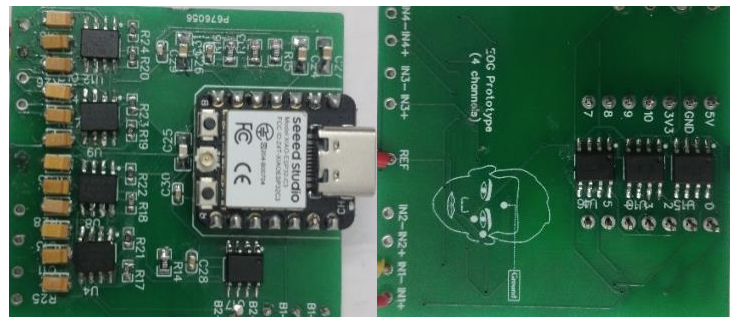

Fig. Assembled Prototype

The signals from the prototype were acquired and successfully plotted and analysed on Arduino IDE, which confirmed the success of the design of the prototype, signal acquisition and the ESP32 utilisation for EOG acquiring signals. These outputs have given confidence to the approach taken & enabled the progress of further stages of development.

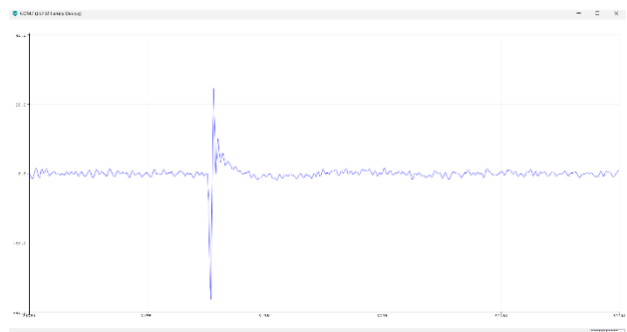

Fig. Real-Time Signal Acquisition (for BLINK)

The next stage is application of Machine Learning algorithms and Micropython for the generation of Text and Voice based outputs using the filtered spectrum. We have reviewed various literature related to Physiological Signal detection using Machine Learning. We have explored various models which can be applied for our use. This phase is a critical phase as we will be creating a dataset with our prototype to train the model for good accuracy.
For analyzing and classifying Electrooculography (EOG) signals using machine learning (ML), various studies have explored methods that focus on detecting and processing eye movement patterns and the spectrum of EOG signals.

1. **Wavelet Transform with Supervised Algorithms:** Wavelet transforms, combined with ML algorithms like K-Nearest Neighbors (KNN), Support Vector Machines (SVM), and Decision Trees (DT), have been used for EOG signal classification, capturing essential frequency components between 0.5 to 50 Hz. For instance, a study using these methods found that SVM performed best with 76.9% accuracy in detecting eye movements such as left, right, up, down, and blink.

2. **LSTM and Statistical Analysis for Blink Detection:** Another approach focused on blink detection and classification by first performing statistical analysis (using first-derivative thresholding) and then applying Long Short-Term Memory (LSTM) networks. This method achieved high accuracy by distinguishing various blink patterns, such as single and double winks, based on slope and peak-trough analysis.

3. **Hybrid Feature Extraction with CNNs:** In a hybrid approach, convolutional neural networks (CNNs) are applied to EOG signals, focusing on different spectral bands to enhance feature extraction. This model, often combined with spectrograms of the signal, has shown to be effective for robust classification across varied movement types, with particular success in applications requiring continuous control, like wheelchair navigation.

| *Study* | *Technique(s)* | *Accuracy* | *Focused Movements* | *Advantages* |
|---|---|---|---|---|
| *Wavelet + SVM/KNN/DT* | Wavelet Transform + SVM/KNN | 76.9%(SVM) | Left, Right, Up, Down, Blink | Good for spectral analysis and spatial classification |
| *LSTM + Statistical* | LSTM + Derivative Analysis | ~91.25% | Blink, Wink | Effective for dynamic movements and blink detection |
| *CNN + Spectral Bands* | CNN + Spectral Bands | ~89% | Various | Strong for real-time applications, e.g., navigation |

**Table-1:** Comparison of Different Models

The ability to perform these steps successfully will lead to the Fully developed product and plum deployment of the communication interface. This final product is intended to make communication for quadriplegic patients easier using effective and reliable communication.

# 6. CONCLUSION

Quadriplegia leads to great communication deficits since it makes voluntary muscle control impossible. In this project, EOG signals are used to develop a communication system based on voice and text for quadriplegic patients to overcome this challenge. This project makes use of electrical potential between the cornea and retina that is employed to determine and track eye movement and convert them into a control system for a computer device system. Using this non-invasive and cheap technique, voice or text output can be produced by impressive four separate eye movements, gaze up or down or to left or right.

This system allows the user to communicate both speech and text in a dynamic manner that is tailored to individual needs. The interface developed in the existing setup has basic communication capabilities which could be improved upon. Further developments could include an increased number of eye movements to communicate more complex messages in the future. An eye blink sensor will be added to the existing system to facilitate its use and simplify system operation. These improvements would improve the user experience of the system and make it more useful and effective for patients with extreme mobility impairments.